\documentclass[useAMS,usenatbib]{mn2e}
\usepackage{times,mathptm}
\usepackage{lscape}
\usepackage{psfig}
\usepackage{subfigure}
\usepackage{multirow}

\usepackage{psfig}
\usepackage{times}
\usepackage{amsmath}
\usepackage{amssymb}
\usepackage{graphicx}
\usepackage{graphics}
\usepackage{rotating}
\usepackage{lscape}
\usepackage{morefloats}
\usepackage{hyperref}
\usepackage{pslatex}

\usepackage{pifont}
\def\apj{ApJ}
\def\apjl{ApJL}
\def\mnras{MNRAS}

\def\araa{ARAA}

\def\apjs{ApJS}

\def\nar{New Astron. Rev.}

\def\gs{\mathrel{\raise0.35ex\hbox{$\scriptstyle >$}\kern-0.6em
\lower0.40ex\hbox{{$\scriptstyle \sim$}}}}
\def\ls{\mathrel{\raise0.35ex\hbox{$\scriptstyle <$}\kern-0.6em
\lower0.40ex\hbox{{$\scriptstyle \sim$}}}}

\def\kms{\,\hbox{km\,s$^{-1}$}}

\def\Wm2{\,\hbox{W}\,\hbox{m}^{-2}}
\def\gsim{\mathrel{\raise0.35ex\hbox{$\scriptstyle >$}\kern-0.6em\lower0.40ex\hbox{{$\scriptstyle \sim$}}}}
\def\lsim{\mathrel{\raise0.35ex\hbox{$\scriptstyle <$}\kern-0.6em\lower0.40ex\hbox{{$\scriptstyle \sim$}}}}
\def\ltsima{$\; \buildrel < \over \sim \;$}
\def\simlt{\lower.5ex\hbox{\ltsima}}
\def\gtsima{$\; \buildrel > \over \sim \;$}
\def\simgt{\lower.5ex\hbox{\gtsima}}

\begin{document}

\title[ALMA resolves star formation in high-$z$ AGN]
{ALMA resolves extended star formation in high-$z$ AGN host galaxies}
\author[C.\ M.\ Harrison et al.]
{ \parbox[h]{\textwidth}{ 
C.~M.~Harrison,$^{\! 1\,\dagger}$
J.~M.~Simpson,$^{\! 1}$
F.~Stanley,$^{\! 1}$
D.~M.~Alexander,$^{\! 1}$
E.~Daddi,$^{\! 2}$
J.~R.~Mullaney,$^{\! 3}$
M.~Pannella,$^{\! 4}$
D.~J.~Rosario,$^{\! 1}$
Ian~Smail$^{\! 1,5}$
}
\vspace*{6pt} \\
$^1${Centre for Extragalactic Astronomy, Department of Physics, Durham University, South Road,
  Durham, DH1 3LE, U.K.}\\
$^2${CEA Saclay, Laboratoire AIM-CNRS-Universit\'e Paris Diderot, Irfu/SAp, Orme des Merisiers, 91191 Gif-sur-Yvette, France}\\
$^3${Department of Physics and Astronomy, University of
  Sheffield, Sheffield, S3 7RH, U.K.}\\
$^4${Universit\"ats-Sternwarte M\"unchen, Scheinerstr. 1, D-81679 M\"unchen, Germany}\\
$^5${Institute for Computational Cosmology, Department of Physics, Durham University, South Road,
  Durham, DH1 3LE, U.K.}\\
$^{\dagger}$Email: c.m.harrison@mail.com \\
}
\maketitle

\begin{abstract}
We present high resolution (0.3$^{\prime\prime}$) Atacama Large Millimeter Array (ALMA)
870\,$\mu$m imaging of five $z$$\approx$1.5--4.5 X-ray detected AGN (with luminosities
of $L_{{\rm 2-8keV}}>10^{42}$\,erg\,s$^{-1}$). These data provide a
$\gtrsim$20$\times$ improvement in spatial resolution over single-dish rest-frame FIR
measurements. The sub-millimetre emission is extended on scales of
FWHM$\approx$0.2$^{\prime\prime}$--0.5$^{\prime\prime}$, corresponding
to physical sizes of 1--3\,kpc (median value of 1.8\,kpc). These sizes
are comparable to the majority of $z$=1--5 sub-millimetre galaxies
(SMGs) with equivalent ALMA measurements. In combination with spectral energy distribution analyses, we
attribute this rest-frame far-infrared (FIR) emission to dust heated by star formation. The implied star-formation rate surface densities are
$\approx$20--200\,$M_{\sun}$\,yr$^{-1}$\,kpc$^{-2}$, which are
consistent with SMGs of comparable FIR luminosities (i.e.,
$L_{\rm IR}$$\approx$[1--5]$\times$10$^{12}$\,L$_{\sun}$). Although limited
by a small sample of AGN, which all have high FIR luminosities, our study suggests that the kpc-scale spatial distribution and
surface density of star formation in high-redshift star-forming galaxies is the same irrespective of the presence of X-ray detected AGN.
\end{abstract}


\section{Introduction}
Understanding the physical processes that drive the growth of
super-massive black holes (SMBHs i.e., active galactic
nuclei; AGN) and how this relates to the growth of their host
galaxies (i.e., star formation), is an ongoing challenge of
observational and theoretical astronomy (e.g., \citealt{Alexander12};
\citealt{Crain15}; \citealt{Volonteri15}). The bulk of star formation and black hole growth
occurred at high redshift (i.e., $z$$\gtrsim$1) and most observational work of high-$z$ galaxies, suggests
that the star-formation rates (SFRs) of AGN hosts are broadly
consistent with the overall star-forming population (e.g.,
\citealt{Stanley15}; \citealt{Azadi15}; \citealt{Banerji15}). However, the {\em
  average} black hole growth rates of high-$z$ massive galaxies do appear to be correlated with
average SFRs (e.g., \citealt{Mullaney12b}; \citealt{Delvecchio14}). These combined results
potentially indicate a common fuelling mechanism for both processes,
but with the AGN activity varying on much shorter timescales than the star
formation (e.g., \citealt{Hickox14}; \citealt{Stanley15}). Unfortunately these
 and similar studies have been limited to spatially-unresolved
 measurements of the star formation, such as those provided by {\em
   Herschel} (e.g., FWHM$\approx$6.5$^{\prime\prime}$ at 100$\mu$m) or SCUBA-2 (e.g., FWHM$\approx$14.5$^{\prime\prime}$ 
 at 850$\mu$m). These measurements hide crucial information on the star formation spatial distribution and surface
densities of star formation.

Arguably the best tracer of star formation in high-$z$ galaxies
is rest-frame far-infrared (FIR) emission
($\lambda$$\approx$8--1000\,$\mu$m). This emission is due to dust that has been heated by young stars inside
star-forming regions (e.g., see \citealt{Lutz14}). For AGN host-galaxies there is some
discussion about the contribution of star formation versus AGN
activity as the source of heating for FIR-emitting dust (e.g., see
\citealt{Hill11}; \citealt{Netzer15}); however, providing direct size
measurements of the emission provides a useful constraint on this issue
(e.g., see \citealt{Lutz15}). In the era of the Atacama Large
Millimeter Array (ALMA), it is now
possible to rapidly build up large samples of high-$z$ galaxies with accurate measurements of the angular sizes of the
rest-frame FIR emission, and consequently to constrain the
spatial distribution and surface density of star formation (e.g.,
\citealt{Simpson15}; \citealt{Ikarashi15};
\citealt{DiazSantos15}). Such work builds on previous sub-arcsecond
  resolution interferometric
continuum observations of a small number highly-selected high-$z$ sub-millimetre
galaxies (SMGs) and AGN (e.g., \citealt{Tacconi06}; \citealt{Clements09}). Comparing the spatial distribution of star
formation in AGN to non-AGN host galaxies of uniformly-selected samples will provide important
information on the feeding and feedback processes involved in SMBH
accretion (e.g., see \citealt{Volonteri15}).

In this letter we present high-resolution (FWHM=0.3$^{\prime\prime}$) ALMA 870\,$\mu$m continuum
measurements of $z$$\approx$1.5--4.5 X-ray identified AGN. This is
based on ALMA data from a programme that was designed to obtain
sensitive SFR measurements (or upper limits) for
X-ray AGN (\citealt{Mullaney15}; \S \ref{sec:methods}). Here, we place constraints on the sizes of
the rest-frame FIR emission in high-$z$ X-ray AGN host galaxies
and hence measure the spatial distribution and surface density of star formation in these sources (\S \ref{sec:extended}). We compare to equivalent ALMA
observations SMGs, to assess if and how
AGN activity in high-$z$ star-forming galaxies is related to SFR surface density (\S \ref{sec:discussion}). Throughout, we adopt $H_0 = 70$\kms\,Mpc$^{-1}$, $\Omega_{\rm{M}} = 0.30$ and $\Omega_{\Lambda}= 0.70$.

\section{Target selection, observations and analysis}
\label{sec:methods}

The data presented in this letter are from a Cycle\,1 Band\,7
ALMA programme to obtain 870\,$\mu$m continuum measurements of
$z$$\ge$1.5 X-ray detected AGN that were selected to be pre-dominantly
faint or undetected in {\em Herschel} measurements.\footnote{This paper makes use of ALMA data:
ADS/JAO.ALMA-2012.1.00869.S. ALMA is a partnership of ESO (representing
its member states), NSF (USA), NINS (Japan), together with NRC
(Canada), NSC and ASIAA (Taiwan), and KASI (Republic of Korea), in
cooperation with the Republic of Chile. The Joint ALMA Observatory is
operated by ESO, AUI/NRAO and NAOJ.} Thirty AGN were targeted that were selected from the {\em Chandra}-Deep Field South (CDF-S;
\citealt{Xue11}), to have X-ray luminosities of $L_{\rm 2-8keV}$$\ge$10$^{42}$\,erg\,s$^{-1}$. The details of how the sample was constructed are
provided in \cite{Mullaney15}; however, we note that they only include
24 sources in their study due to specific constraints on the redshifts and stellar
masses. Here we exploit {\em all} of the high signal-to-noise ratio (SNR)
detections from these data, including serendipitous detections (see
below).

\subsection{ALMA observations}

The 30 primary targets were split into three groups
containing 7, 11 and 12 targets each. The first two groups were
observed in two observing blocks, whilst the third group was observed once. The array
configuration contained 26 ALMA antennae, with a maximum baseline of 1300\,m and
median baselines of $\approx$200\,m. The observations are sensitive to
a maximum angular scale of 4--6$^{\prime\prime}$, at which we expect to
recover all of the rest-frame FIR-emission (see \citealt{Simpson15,Ikarashi15}).  Each target was observed using 7.5\,GHz
of bandwidth, centred on 351\,GHz (i.e., $\approx$870\,$\mu$m), with
on-source exposure times of 2.5--7\,min. All measurement sets have a full
compliment of calibrator observations (amplitude, phase and
bandpass). Full details of the observations and data reduction will be
presented in Stanley et~al. (in prep). 

\subsection{Data reduction and source detection}
\label{sec:sources}

\begin{figure}
\centerline{
\psfig{figure=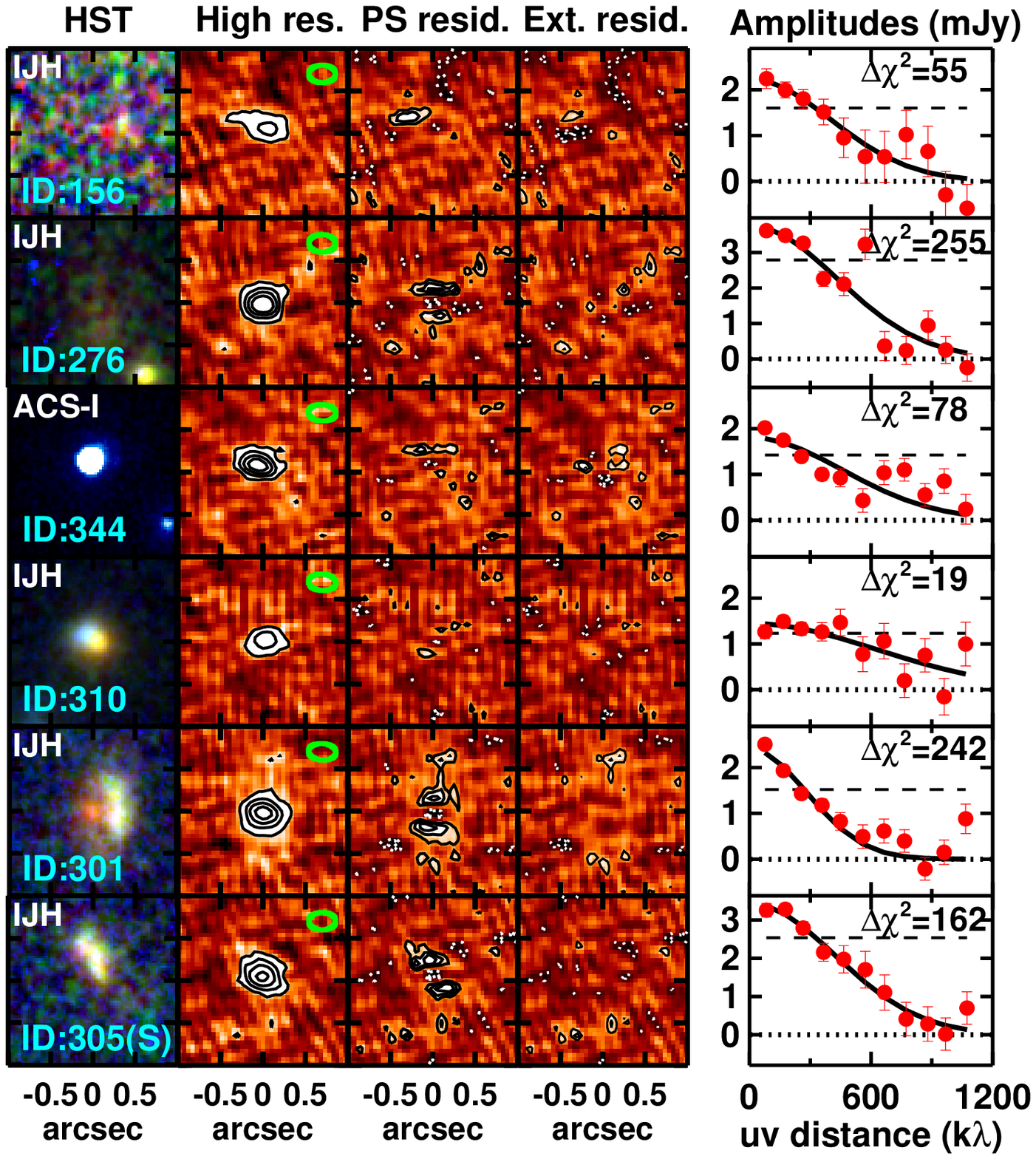,width=\columnwidth,angle=0}
}
\caption{{\em HST} and ALMA data for our six high SNR detections,
  comprising of five X-ray AGN and one serendipitous source
  (ID:305(S); \S \ref{sec:sources}). From left to right: (1) {\em
    HST} images (see \S
  \ref{sec:multi}); (2) ALMA high-resolution
image (contours at [3,7,11,15]$\times\sigma$, the green ellipses
illustrate the synthesised beam); (3) residual image from a point
source fit (contours at [$-4$,$-3$,$-2$,2,3,4]$\times\sigma$); 
(4) residual image from an elliptical Gaussian fit (contours as in
(3)); (5) real components of the visibilities (mJy). All images are 2$^{\prime\prime}$$\times$2$^{\prime\prime}$ and are
centred on the ALMA detection position. In the final panel the dashed lines are the best fit constant amplitude models
(i.e., ``point source" models) and the solid curves are the Gaussian models (i.e., ``extended" models). The $\Delta\chi^{2}$ values
are the differences between $\chi^{2}$ for these two fits and
indicate strong evidence to favour extended structure (see \S \ref{sec:extended}).}
\label{fig:maps}
\end{figure}

The data were processed using the {\sc Common Astronomy Software
  Application} ({\sc casa}; version 4.4.0; \citealt{McMullin07}) and
imaged using the {\sc clean} routine provided by {\sc
  casa}. We used the most recent version of the ALMA data reduction
pipeline to calibrate the raw data. However, the calibrated data were then visually inspected and,
where appropriate, we repeated the pipeline calibration including additional
data flagging. To image the data, we largely follow the methods described in detail in
\cite{Simpson15} and so we only provide brief details here. We created
two sets of images: (1) ``detection images'' (FWHM$\approx$0.8$^{\prime\prime}$) and  (2)
``high-resolution images'' (FWHM$\approx$0.3$^{\prime\prime}$). For both sets of images we initially
create ``dirty'' images and identify emission detected with
SNR$\ge$5. We then place a tight clean mask around the emission from
the source and iteratively clean down to 1.5$\sigma$ within these
regions. Finally, we measure the noise in the cleaned image, and repeat
the cleaning process around SNR$\ge$4 sources.

For the detection images, we applied natural weighting and a Gaussian
taper, resulting in synthesised beams of
(0.8$^{\prime\prime}$--0.9$^{\prime\prime}$)$\times$0.7$^{\prime\prime}$. The noise of the final cleaned images have a range of
$\sigma_{870}$=0.10--0.25\,mJy\,beam$^{-1}$. To create the
high-resolution images we used Briggs weighting (robust parameter =
0.5) to obtain synthesised beams of (0.3$^{\prime\prime}$--0.4$^{\prime\prime}$)$\times$0.2$^{\prime\prime}$. These
final cleaned images have a noise of $\sigma_{870}$=0.07--0.18\,mJy\,beam$^{-1}$.

We searched for ALMA sources that are detected
within the primary beam of the high-resolution images with {\em peak} SNRs$\gtrsim$9. Above this detection threshold we can make
measurements of the continuum sizes in these images and compare
directly to the SMGs with equivalent measurements in
  \cite{Simpson15} (see \S \ref{sec:discussion}). Across all of the images we
obtain six sources, with peak SNRs$\gtrsim$9, of which three are primary X-ray
AGN targets for this programme (\citealt{Mullaney15}), two are
serendipitous X-ray AGN and one is a serendipitous source which is not
X-ray detected (Table~1). This low ALMA detection rate of the primary
AGN targets is driven by a selection which prioritised AGN with low
{\em Herschel} FIR fluxes and a discussion of this is provided in
\cite{Mullaney15}. In Section~\ref{sec:discussion} we compare
our detected sources to SMGs with similar fluxes and luminosities. The ALMA detection images and high-resolution
images for these six targets are shown in Figure~\ref{fig:maps}. Peak
flux densities are measured directly from the images (calibrated in
units of Jy\,beam$^{-1}$). Total flux densities are measured
using high-resolution images that are converted to Jy\,pixel$^{-1}$
and 1$^{\prime\prime}$ diameter apertures. We note that we obtain consistent
flux measurements if we use the detection images (i.e., agreement within 20\% in all
cases and a median ratio between measurements of 1.0). Uncertainties are calculated by taking the 1$\sigma$
  distribution from placing hundreds of random apertures across the images.  

\begin{table}
\begin{center}
{\footnotesize
{\centerline {\sc Properties of the high SNR ALMA-detected X-ray AGN}}
\begin{tabular}{cccccc}
\hline
ID &      $z$ & $\log L_{X}$        & $S_{870\mu m}$  & $\log L_{{\rm IR,SF}}$   & FWHM\\
     &         & (erg\,s$^{-1}$) & (mJy)        &(L$_{\sun}$) & (arc sec) \\
\hline
156 & 4.7$_{-1.9}^{+1.2}$             &  43.6     &   2.3$\pm$0.4   &     12.3$_{-0.3}^{+0.2}$    &   0.49$\pm$0.11            \\
276 & 1.52$_{-0.16}^{+1.57}$             &  42.1     &   3.7$\pm$0.3  &      12.6$_{-0.4}^{+0.2}$   &   0.20$\pm$0.03             \\
301 & 2.47$_{-0.26}^{+0.06}$             &  43.3     &   2.70$\pm$0.19   &       12.4$_{-0.2}^{+0.1}$        &   0.26$\pm$0.04             \\
310 & 2.39$_{-0.23}^{+0.09}$             &  43.2     &   1.44$\pm$0.28   &       12.1$_{-0.3}^{+0.1}$        &   0.23$\pm$0.06            \\
344 & 1.617                                   &  43.4    &2.02$\pm$0.19  &        12.3$_{-0.1}^{+0.2}$         &    0.17$\pm$0.05           \\
305(S)$^{\star}$ & 2.93$_{-0.10}^{+0.10}$         &  -          &3.6$\pm$0.3 &        12.1$_{-0.3}^{+0.1}$         &    0.32$\pm$0.04         \\
\hline
\hline
\label{tab:targets}
\end{tabular}
}
\vspace{-0.8cm}
\caption{X-ray ID (\citealt{Xue11}); redshift (see
  \S \ref{sec:multi}); 2--8\,keV X-ray luminosity; 870\,$\mu$m
  galaxy-integrated primary-beam corrected flux density; FIR luminosity from star formation (see
  \S \ref{sec:multi}); de-convolved 870\,$\mu$m FWHM (major axis; see
  \S \ref{sec:extended}). $^{\star}$ The final target in the
  table is a serendipitous detection in the ALMA
  map centred on the X-ray source, ID:305.}
\end{center}
\end{table}

\subsection{Multi-wavelength properties}
\label{sec:multi}

\begin{figure}
\centerline{
\psfig{figure=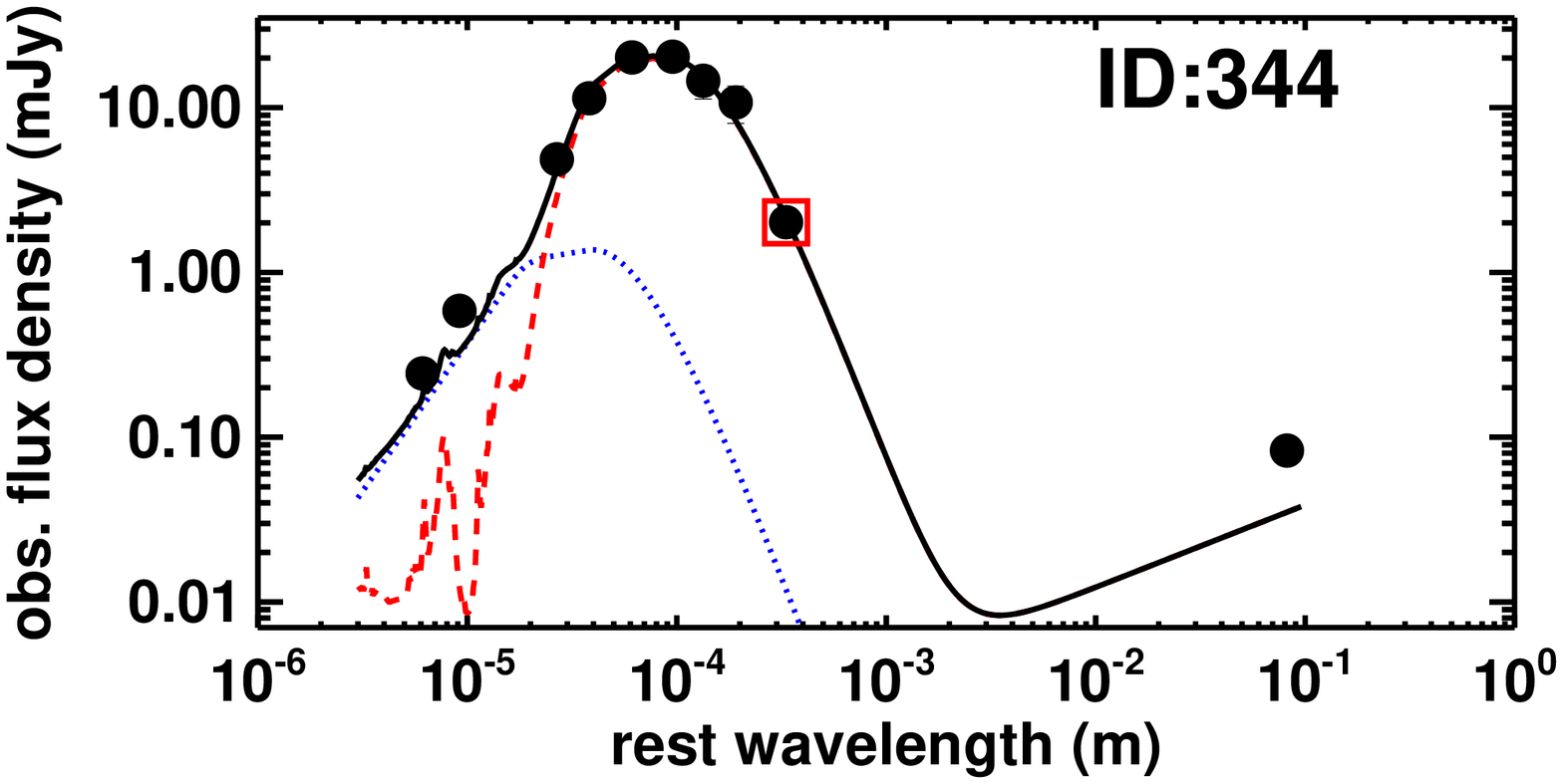,width=0.48\textwidth,angle=90}
}
\caption{Example infrared SED for one of our targets. The
  data are from {\em Spitzer}, {\em Herschel}, ALMA and the Very Large Array
  (ALMA point highlighted with an open square) and the solid curve shows the
  best fit model (see \S \ref{sec:multi}). The uncertainties are
  smaller than the symbol size. The data are fit with an AGN
  template (dotted curve) and star-formation template (dashed
  curve). FIR luminosities ($L_{\rm IR,SF}$) are derived by
  integrating the star-formation contribution only,
  from 8--1000\,$\mu$m (see Table~1). The radio data point
  is not included in the fit and may be slightly higher than the
  star-forming template due to an AGN contribution to the radio emission.}
\label{fig:sed}
\end{figure}

The details of our final six targets are tabulated in
Table~1. For the AGN we adopt the photometric redshifts and 1$\sigma$ uncertainties provided in
\cite{Hsu14}. They identified the optical counterparts of the X-ray sources in CDF-S and performed
detailed spectral energy distribution (SED) analyses to optical
through mid-infrared photometric data, including host galaxy
and AGN templates. We adopt the available spectroscopic redshift for
ID:344 (\citealt{Szokoly04}) and for the serendipitous target we use
the photometric redshift and 1$\sigma$ uncertainty from the
3D-{\em HST} team (\citealt{Skelton14}). We use these redshifts and the
2--8\,keV X-ray fluxes from \cite{Xue11} to calculate X-ray
luminosities, assuming a power-law index of $\Gamma$=1.4 (Table~1). To determine the position of the ALMA sources with respect to the
optical emission, we collated the $I$-band, and where
possible, $J$ and $H$ band {\em HST} observations of our targets
(Fig.~\ref{fig:maps}; \citealt{Guo13}). 

We derive total infrared luminosities ($\lambda$=8--1000\,$\mu$m) by
fitting SEDs to the available {\em Spitzer}, deblended {\em
  Herschel}-PACS; deblended {\em
  Herschel}-SPIRE and ALMA photometry (i.e., $\lambda$=16\,$\mu$m--870\,$\mu$m; see Fig.~\ref{fig:sed}). The details
of the SED fitting routine and the compilation of the
non-ALMA photometry are detailed in \cite{Stanley15}. Briefly, the
fitting routine finds the best fit SED from normalising various combinations of empirical star-formation templates and an
AGN template, taking into account photometric data points,
uncertainties, and upper limits. Using the best-fit SEDs we derived total infrared
luminosities, $L_{{\rm IR,SF}}$, due to star formation only (i.e., subtracting off any
identified AGN contribution; see Fig.~\ref{fig:sed}). In two cases (ID:156 and
ID:276) there are no {\em Spitzer} or {\em Herschel} detections and
therefore we use the ALMA measurement only. We believe that
the ALMA photometry is well described by emission due to star formation
because, based on a range of AGN templates (e.g., see \citealt{Netzer15}), there would be a bright {\em Spitzer} 24\,$\mu$m
detection if it was AGN dominated. Furthermore, in
the other four cases the SEDs indicate that the ALMA photometry
is dominated by star formation (e.g., Fig~\ref{fig:sed}; Stanley et~al. in prep). Finally, we assess if the sub-mm
fluxes could have a contribution from radio synchrotron emission. Only
three of our sources (156, 344 and 305[S]) are detected in the deep 1.4\,GHz radio imaging
of \cite{Miller13} (typical sensitivity of 7.4\,$\mu$Jy per
2.8$^{\prime\prime}$ by 1.6$^{\prime\prime}$ beam). The flux densities are
88\,$\mu$Jy, 83\,$\mu$Jy and 41\,$\mu$Jy, respectively, which are over an
order of magnitude lower than the ALMA flux densities (e.g., Fig.~\ref{fig:sed}). We conclude that the
ALMA photometry has negligible contribution from synchrotron emission.

\section{Extended far-infrared emission}
\label{sec:extended}

We have compiled a sample of five X-ray AGN, and one serendipitous
target, with high SNR ALMA 870\,$\mu$m continuum detections in our
high-spatial resolution images
(FWHM$\approx$0.3$^{\prime\prime}$). For the
redshifts of our targets these data cover rest-frame far-infrared
wavelengths of $\approx$150--330\,$\mu$m. In this section we assess if this
emission is extended and measure intrinsic (deconvolved) sizes. We are
specifically interested in comparing to the redshift-matched ($z$=1--5) SMGs with ALMA 870\,$\mu$m sizes presented in
\cite{Simpson15}. These ALMA observations were taken at the same resolution
as those presented here (i.e., FWHM$\approx$0.3$^{\prime\prime}$) 
and we can derive directly comparable size measurements. We also
refer briefly to \cite{Ikarashi15} who make size measurements of
SMGs; however, they focus on higher redshift sources, typically have
lower spatial resolution data and employ different methods to this study.

\begin{figure*}
\centerline{
\psfig{figure=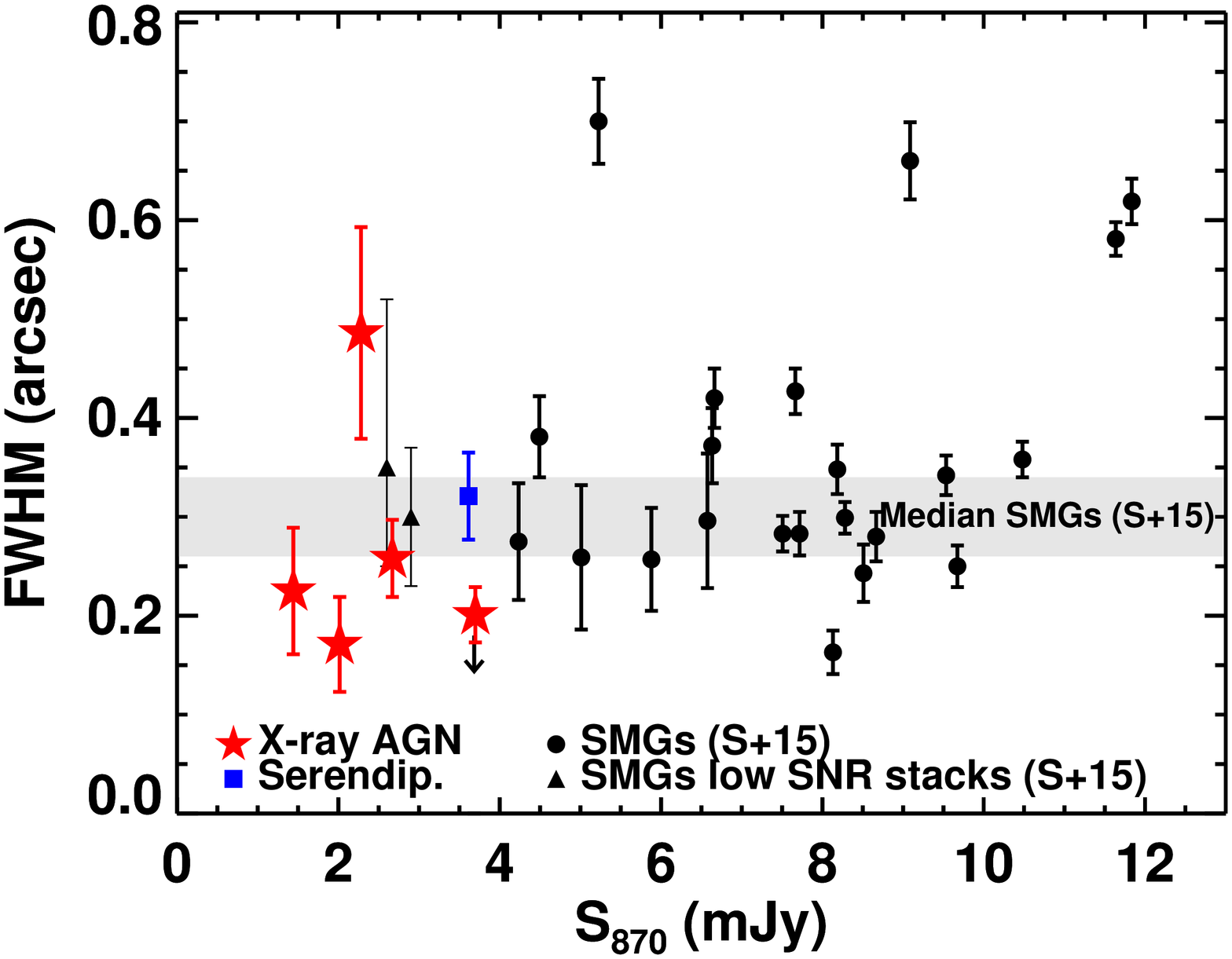,width=0.45\textwidth,angle=90}
\psfig{figure=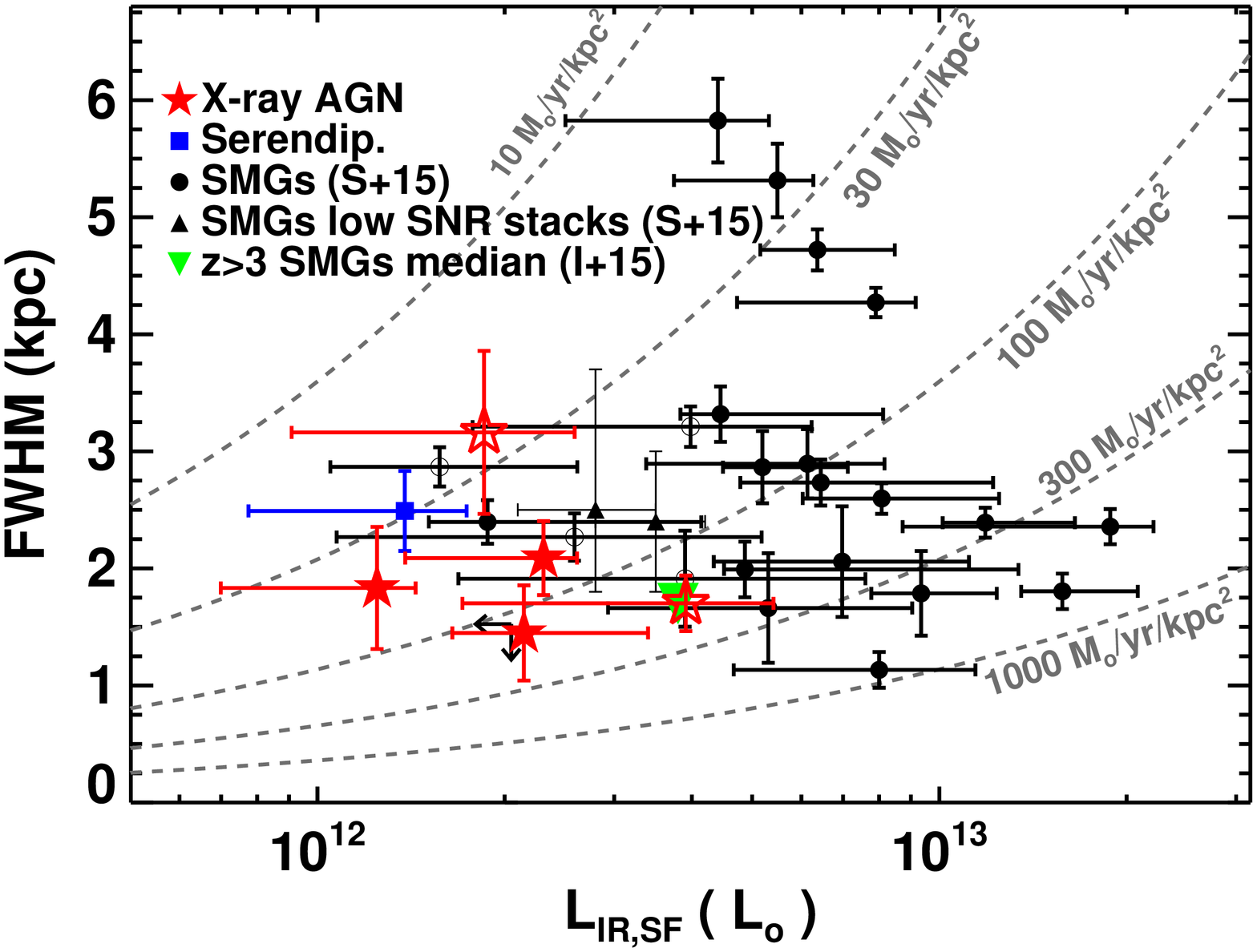,width=0.45\textwidth,angle=90}
}
\caption{{\em Left:} Intrinsic angular size of the 870\,$\mu$m emission
  as a function of flux density for our high-$z$ X-ray AGN, the
  serendipitous detection and high-$z$ SMGs (\protect\citealt{Simpson15};
  ``S$+$15''). The shaded region shows their median value plus/minus
  1$\sigma$. {\em Right:} Physical size as a function of FIR
  luminosity (see \S~\ref{sec:multi}). We also show the median value
  of the long-baseline ALMA observations from \protect\cite{Ikarashi15} (``I$+$15''). The hollow symbols correspond to single-band (870\,$\mu$m) derived infrared
luminosities. The dashed curves show constant values of SFR surface
density.}
\label{fig:sizes}
\end{figure*}

We make use of both the raw visibility data and our cleaned images to
search for extended continuum emission in the ALMA data and measure
sizes for our six sources. Firstly, we explore the raw visibilities by assessing how
the amplitude of the data change as a function of $uv$
distance.  Increasing $uv$ distance corresponds to smaller angular
scales. Therefore, a point source has a constant amplitude
across all $uv$ distances and an extended source has a {\em decreasing} amplitude
as a function of increasing $uv$ distance (see e.g.,
\citealt{Rohlfs96}). For each target we align the phase centre of the
data to the position of the ALMA source and then extract the
visibility amplitudes, binning across the $uv$ distance in steps of
100k$\lambda$, and calculating the error on the mean in each bin (see Fig.~\ref{fig:maps}).\footnote{We
note that our targets do not have other bright ALMA sources in
close proximity (which may complicate the analysis of the $uv$
data). However, we obtain consistent results if we model and subtract all other
5$\sigma$ sources before extracting the visibilities.} We model the amplitude-$uv$ data with: (1) a
constant amplitude (applicable for point source emission) and (2) a Gaussian
(applicable for a Gaussian distribution of emission). 

The amplitude $uv$ data and our fits are shown in
Figure~\ref{fig:maps}. In all six cases the amplitudes are
better described as decreasing, rather than constant, with $uv$ distance. The $\Delta\chi^{2}$ values between the two fits range from
19--255. Using the Bayesian Information Criterion (\citealt{Schwarz78}), which takes
into account the number of parameters in each fit, these values indicate strong evidence in favour of the Gaussian model (e.g., \citealt{Mukherjee98}).

To measure the intrinsic far-infrared sizes of our targets we follow
the same methods as those presented in \cite{Simpson15}. That is, we use the {\sc imfit} routine in {\sc casa} to fit an elliptical
Gaussian model (convolved with the synthesised beam) to the 870\,$\mu$m
emission in our high-resolution images (see \citealt{Simpson15} and
\citealt{Ikarashi15} for various tests of this routine). In all cases {\sc
  imfit} returns spatially-resolved fits, in agreement with our
conclusions based on the $uv$ data above. The fits are a good description of
the data and show reduced residuals compared to point source model
fits, as shown in Figure~\ref{fig:maps}. We quote the sizes (major axes) and uncertainties
returned by {\sc casa} in Table~1 and plot them in
Figure~\ref{fig:sizes}. For comparison, we derive sizes from 
Gaussian fits to the $uv$-amplitude data (Fig.~\ref{fig:maps}; following
e.g.,~\citealt{Rohlfs96}). These values assume symmetry and will be systematically low compared to elliptical fits of the images (see
\citealt{Ikarashi15}); however, these $uv$ sizes agree within 1--3$\sigma$ of the image-derived
sizes, with a median ratio of FWHM$^{uv}$/FWHM$^{{\rm image}}$=0.9, in
agreement with that found by \cite{Simpson15} for SMGs. 

We perform two further tests to verify there is extended emission in our six sources and that our size measurements are
reliable. Firstly, we compare the peak and galaxy-integrated fluxes in
the high-resolution images. We find $S^{0.3}_{\rm pk}$/$S^{0.3}_{\rm
  int}$=0.49--0.74. These ratios indicate extended structure since a point source
would have a peak flux density equal to the total flux
density. Finally, we compare the peak flux densities in the detection
images to those in the high-resolution images and obtain
ratios of $S^{0.3}_{\rm pk}$/$S^{0.8}_{\rm pk}$=0.56--0.82. This drop
in flux is strong evidence for emission that is more resolved at
higher resolution. These ratios agree within 1--20\% of the predicted values we obtain by taking out best {\sc imfit} models and
convolving them with the appropriate beams. This places further
confidence on our size measurements described above. 

\section{Discussion}
\label{sec:discussion}
We have identified extended 870\,$\mu$m emission in the $z\approx$1.5--4.5 host galaxies of
five X-ray AGN (see Figure~\ref{fig:maps}). The measured intrinsic (i.e., deconvolved)
sizes are FWHM=0.2$^{\prime\prime}$--0.5$^{\prime\prime}$ and correspond to projected physical sizes of
$\approx$1--3\,kpc, with a median value of 1.8\,kpc (see Fig.~\ref{fig:sizes}). We have used infrared
SEDs to show that our ALMA photometry is consistent with star-forming SEDs (e.g., Fig.~\ref{fig:sed}) and consequently
we attribute this rest-frame FIR emission (i.e., $\approx$150--330\,$\mu$m) to 
dust heated by star formation. Furthermore, the
observed spatial extent of the FIR emission on $>$1\,kpc scales is
challenging to explain with dust that is heated directly by AGN without any contribution from star formation.

In Figure~\ref{fig:sizes} we compare our FIR size measurements with the
$z$$\approx$1--5 SMGs from the UDS field that were observed with ALMA
by \cite{Simpson15} and one serendipitous star-forming galaxy from our
data. These sources have a similar redshift range to our
X-ray AGN and have available FIR luminosities that are calculated
following similar infrared SED analyses to those we applied to our
AGN. Crucially, the size measurements are obtained from
equivalent observational data sets and by using the same techniques as
applied here. \cite{Simpson15} present results for 23
SMGs for which they have high SNR ALMA detections, and hence reliable
size measurements. They stack the data for a further 25 sources which
have lower SNR detections (see Fig.~\ref{fig:sizes}). These SMGs are not
identified as X-ray AGN based on the X-ray coverage of this field
(\citealt{Ueda08}). This X-ray data coverage is relatively shallow;
however, very deep studies in CDF-S find the X-ray AGN detection rate of
ALMA-identified SMGs to be $\approx$20\% (\citealt{Wang13}) and will make a
minor contribution to the overall SMG sample. 

The median intrinsic FIR size of the SMGs is FWHM$=$0.3$^{\prime\prime}$$\pm0.04^{\prime\prime}$, with a corresponding
median physical size of 2.4$\pm$0.2\,kpc
(\citealt{Simpson15}). Therefore, the sizes of the rest-frame
FIR emission of our X-ray AGN host galaxies, are consistent with the
typical sizes of SMGs (see Fig.~\ref{fig:sizes}). Four of the most luminous SMGs (with $L_{{\rm IR,SF}}>4\times10^{12}$L$_{\sun}$) have very large sizes of
4--6\,kpc. Our sample of X-ray AGN do not reach these
high FIR luminosities and due to the low number of targets, we cannot
conclude anything significant about the lack of very large sizes in our
AGN sample. 

To derive SFR surface densities we follow
\cite{Simpson15}. That is, we convert infrared luminosities to SFRs, following \cite{Kennicutt98}
(converting to a Chabrier IMF), and assume a uniform surface
density with a radius of FWHM/2 (see tracks in
Fig.~\ref{fig:sizes}). We obtain SFR surface densities of
$\approx$20--200\,$M_{\sun}$\,yr$^{-1}$\,kpc$^{-2}$, similar to the
subset of SMGs with comparable FIR luminosities to our sample (Fig.~1). These results provide
evidence that the kpc-scale star formation distribution and surface densities of
high-$z$ star-forming galaxies are independent of the presence of
an X-ray AGN. This implies that the physical mechanisms
driving the star formation (see discussion in \citealt{Simpson15} and
\citealt{Ikarashi15}) are similar in these two populations. We note that our AGN have SFRs of
$\approx$130--400\,M$_{\sun}$\,yr$^{-1}$, stellar masses of $\approx$(2--20)$\times$10$^{10}$\,M$_{\sun}$ and corresponding
specific SFRs of $\approx$1--20\,Gyr$^{-1}$ (see \citealt{Mullaney15}). These values
are similar to SMGs of comparable luminosity (e.g.,
\citealt{Simpson14}); however, they may represent the high
end of the (s)SFR distribution of X-ray AGN (see \citealt{Mullaney15}). 

Based on {\em Herschel}-160\,$\mu$m imaging, $z<0.05$
X-ray AGN typically have larger rest-frame FIR sizes than our high-$z$
samples, reaching sizes of FWHM$\approx$5--30\,kpc (\citealt{Mushotzky14}). However, we caution that
these results are based on low-resolution data, leading to some
sources with upper limits on the measured sizes. These local AGN
have a wide range of SFR surface densities, covering $\gtrsim$2.5\,dex, with
the majority of values being low (i.e., $<$0.1\,M$_{\sun}$\,yr$^{-1}$\,kpc$^{-2}$) compared to our high-$z$ X-ray
AGN. These local AGN have similar X-ray luminosities
to our sample (a proxy for black-hole accretion rate),
but typically much lower FIR luminosities (a proxy for SFR; i.e.,
$L_{\rm IR}$$\approx$10$^{9}$--10$^{11}$\,L$_{\sun}$;
\citealt{Shimizu15}). Therefore, the star
formation sizes and surface densities appear to be insensitive to the
presence of an X-ray AGN. \cite{Lutz15} recently reached a similar
conclusion for local galaxies that host optical AGN using
{\em Herschel} data. In contrast, extreme SFR
surface densities are only associated with the most extreme star-forming galaxies (Fig.~\ref{fig:sizes}), which may provide evidence
for different fuelling mechanisms (e.g., see \citealt{Daddi10}). We
note that all of these measurements hide information of $\lesssim$1\,kpc structures.

Although this study is limited to sources with FIR luminosities of
$L_{\rm IR,SF}\approx$[1--5]$\times$10$^{12}$\,L$_{\sun}$, we now have
a first-order assessment of the size scale of the star formation that
has been measured for high-$z$ X-ray AGN by {\em Herschel} studies
(e.g., \citealt{Stanley15}). These {\em Herschel} studies have shown
that the average SFRs appear to be broadly independent of AGN luminosity, which trace instantaneous black-hole
accretion rates. These results may not be surprising because the star
formation appears to be occurring on scales of a few kpc,
orders of magnitude larger than the immediate vicinity of the black
hole. Indeed, local studies have shown a tighter correlation between
nuclear SFRs, compared to $\gtrsim$1\,kpc-scale SFRs, and black hole accretion rates
(\citealt{DiamondStanic12b}). By collating large samples of spatially-resolved ALMA data
of high-$z$ galaxies, it will be possible to assess the relationship
between SFR surface density and AGN activity, providing fundamental insight into feeding and feedback
mechanisms governing galaxy and black hole growth.

\vspace{-0.5cm}
\subsection*{Acknowledgements}
We thank the referee for constructive comments. We acknowledge the Science and Technology Facilities Council
(CMH; DMA; DJR; IRS; grant code ST/L00075X/1). IRS and JMS acknowledge the ERC (Dustygal
321334) and IRS acknowledges a RS/Wolfson Award. FS acknowledges a Durham Doctoral Scholarship.

\bibliographystyle{mn2e}

 
\end{document}